\documentclass[12pt]{article}

\usepackage{amsmath}
\usepackage{amssymb}
\usepackage{amsfonts}
\usepackage{graphicx}
\begin{document}

\begin{center}
{\large \bf{ CMB two-point angular correlation function in the Ellipsoidal Universe}}
\end{center}

\vspace*{1.0 cm}

\begin{center}
{
Paolo Cea~\protect\footnote{Electronic address:
{\tt paolo.cea@ba.infn.it}}  \\[0.5cm]
{\em INFN - Sezione di Bari, Via Amendola 173 - 70126 Bari,
Italy} }
\end{center}

\vspace*{1.0 cm}

\begin{abstract}
\noindent 
We suggest that the Ellipsoidal Universe cosmological model, proposed several years ago  to account for the low quadrupole
temperature-temperature correlation of the Cosmic Microwave Background, can also provide 
temperature-temperature two-point angular correlation function in reasonable agreement with Planck observations.
\end{abstract}

\vspace*{0.6cm}

\vspace{0.2cm}
\newpage
\noindent
\section{Introduction}
\label{s-1}
The latest release of the Cosmic Microwave Background (CMB) anisotropy data by the Planck Collaboration has confirmed
the $\Lambda$ Cold Dark Matter ($\Lambda$CDM) cosmological model to an unprecedented level of statistical
significance (Planck results can be obtained by means of the Planck Legacy Archive~\cite{Planck:2018}).
However, at large angular scales there are several anomalous features in the temperature maps, such as the alignment of
multipoles and the hemispherical power asymmetry. The most notable discrepancy resides in the low quadrupole moment
which, indeed, signals an important suppression of power at large scales.
In the standard CMB analysis the temperature fluctuations are expanded in spherical harmonic:
\begin{equation}
\label{1}
\frac{\Delta T(\vec{n})}{T_0} \;  = \; 
\sum_{\ell, m }  \; a_{\ell m} \, Y_{\ell m}(\theta,\phi)  \; 
\end{equation}
where $\theta,\phi$ are the polar angles of the unit vector $\vec{n}$, and
\begin{equation}
\label{2}
T_0 \;  \simeq   \; 2.7255 \; K 
\end{equation}
is the actual average temperature of the CMB radiation~\cite{Fixen:2009}. The properties of the CMB anisotropy are fully 
characterised by the angular power spectrum:
\begin{equation}
\label{3}
C_{\ell} \; = \; \frac{1}{2 \ell+1} \sum_{m=-\ell}^{+\ell} \,  \left <  | a_{\ell m} |^2 \right > \; .
\end{equation}
where the brackets denote the full-sky average. Indeed, assuming statistical isotropy and gaussianity, i.e. the coefficients $a_{\ell m}$ are
independent gaussian random variables of zero mean, the parameter $C_{\ell}$ turns out to be  the best estimator. Considering that there is only
one sky there is an intrinsic uncertainty in the knowledge of $C_{\ell}$ given by the cosmic variance:
\begin{equation}
\label{4}
\sigma^{CV}_{\ell} \; = \; \sqrt{\frac{2}{2 \ell+1}} \; C_{\ell}   \; .
\end{equation}
The quadrupole anisotropy refers to the multipole $\ell = 2$. Defining
\begin{equation}
\label{5}
( \Delta T_{\ell} )^2  \;  =   \;  T_0^2 \; \frac{\ell(\ell+1)}{2 \pi}  \; C_{\ell} \; \; , \; \; 
\end{equation}
the observed quadrupole anisotropy~\cite{Planck:2018}
\begin{equation}
\label{6}
( \Delta T_{2} )^2  \;  \simeq \;  225.9 \;  \;  \mu \, K^2 \;  , 
\end{equation}
turns out to be much smaller than the quadrupole anisotropy expected  according to the  best-fit  $\Lambda$CDM model to the Planck 2018 data~\cite{Planck:2018}:
\begin{equation}
\label{7}
( \Delta T^{\Lambda CDM}_{2} )^2  \;  \simeq  \;  1150 \;   \mu \, K^2 \; .
\end{equation}
The quadrupole anisotropy is affected by the largest uncertainty due to the cosmic variance, so that the best-fit $\Lambda$CDM quadrupole
anisotropy Eq.~(\ref{7}) differs from the observed value Eq.~(\ref{6}) by less than two standard deviations. As a consequence, it could well be
that the quadrupole anomaly is due to a mere statistical fluctuation.  \\
Nevertheless, several years ago in Refs.~\cite{Campanelli:2006,Campanelli:2007}  to account for the observed suppression of power 
in the quadrupole temperature anisotropy it was  proposed the Ellipsoidal Universe cosmological model.   
In the Ellipsoidal Universe model the flat   Friedmann-Lema\^itre-Robertson-Walker metric is replaced by the following Bianchi I 
anisotropic metric with planar symmetry:
\begin{equation}
\label{8}
d s^2 \; = \; - c^2 \; dt^2 \; + \;  a^2(t) \; \left ( \delta_{ij} \; - \;  e^2(t) \; n_i \; n_j  \right )  dx^i  \; dx^j \; 
\end{equation}
where $e(t)$ is the ellipticity and the unit vector $\vec{n}$ determines the direction of the planar symmetry axis. More precisely, in  Refs.~\cite{Campanelli:2006,Campanelli:2007} it was assumed that the temperature fluctuations satisfied:
\begin{equation}
\label{9}
\Delta T   \;  \simeq  \;  \Delta T^I   \;  + \; \Delta T^A   \;\;  
\end{equation}
where $\Delta T^I$ and  $\Delta T^A$ were the temperature fluctuations induced by the cosmological scalar perturbations and
by the spatial  anisotropy of the metric. After that, the contributions to the temperature fluctuations due to the metric anisotropy were
estimated by simple geometric arguments or, equivalently, by means of the integrated Sachs-Wolf effect. In this way it was suggested
that a small ellipticity at decoupling could explain both the almost planarity and the suppression of power of the quadrupole moment.
Subsequently, in a series of papers~\cite{Cea:2010,Cea:2014,Cea:2020} we solved at large scales the Boltzmann equation for the
photon distribution functions by taking into account the effects of the inflation produced primordial scalar perturbations and
the anisotropy of the geometry. We showed that, in fact, at large scales one recovers Eq.~(\ref{9}). We, also, showed that 
the anisotropy of the spatial geometry contributes mainly to the temperature quadrupole anisotropy without affecting the higher
multipoles since:
\begin{equation}
\label{10}
\ell \; (\ell \; +\; 1) \; C^A_{\ell} \; \sim\;  \frac{1}{ \ell}   \; \; \; , \; \; \; \ell \; \gtrsim 3 \; .
\end{equation}
 Moreover, we found that the ellipsoidal geometry of the universe induces sizeable polarisation signal only at large scales ($\ell \, \lesssim \, 10$)
 without invoking the reionization processes. Finally, in Ref.~\cite{Cea:2020} we were able to fix the eccentricity at decoupling and the
 polar angles $\theta_n , \phi_n$ of the direction of the symmetry axis  $\vec{n}$ such that the quadrupole temperature-temperature
 correlation matched exactly the Planck 2018 value, Eq.~(\ref{6}) obtaining:
\begin{equation}
\label{11}
 e_{dec}  \;   =  \;  8.32 \; \pm \; 1.32 \; \; 10^{-3}    \;   \; ,
\; \;  \theta_n  \;   \simeq  \;  73^\circ    \;  \;  , \; \;  \phi_n \;  \simeq   264^\circ  \;  \; .
\end{equation}
Another notable large-scale anomaly  was displayed by the CMB temperature-temperature correlation function.
In fact, it is now well established that at large angular scales the temperature two-point angular correlation function is found to 
be smaller than expected within the $\Lambda$CDM cosmological model.
The main aim of the present paper is to show that the anomalies displayed by the two-point angular correlation function could be
solved by the Ellipsoidal Universe cosmological model. \\
The remaining part of the paper is organised as follows. In Sect.~\ref{s-2}
we critical discuss  the large scale anomalies of the two-point angular correlation function focusing, for definiteness,  on  the Planck 2013
 and Planck 2015 data. Sect.~\ref{s-3} is devoted to the two-point temperature correlation function within the Ellipsoidal Universe model.
Finally, in Sect.~\ref{s-4} we, briefly, summarise the results presented in this paper and draw our conclusions.
\section{Two-point temperature-temperature angular correlation function}
\label{s-2}
\begin{figure}[t]
\vspace{-1.0cm}
\includegraphics[width=0.5\textwidth,clip]{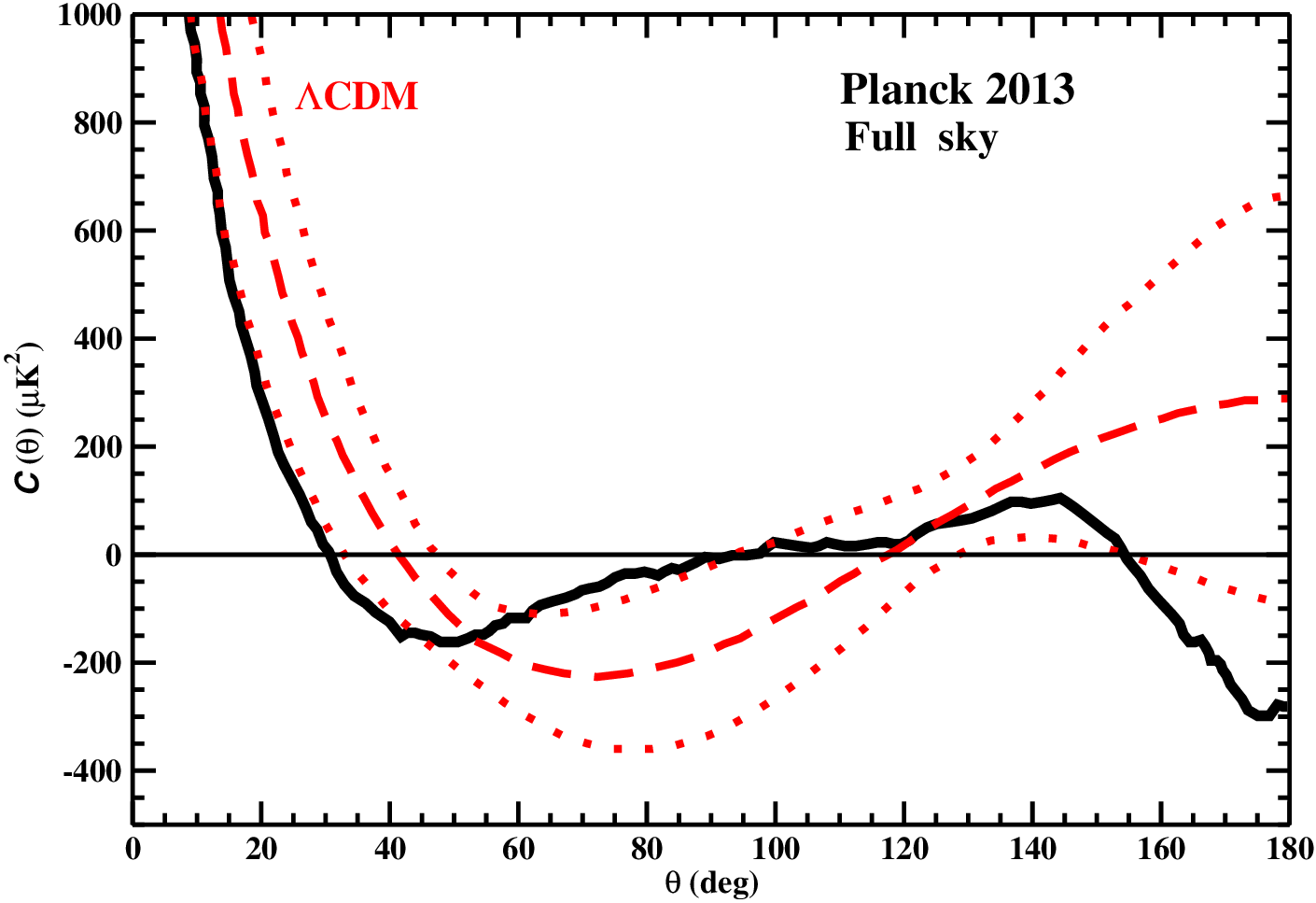}
\includegraphics[width=0.5\textwidth,clip]{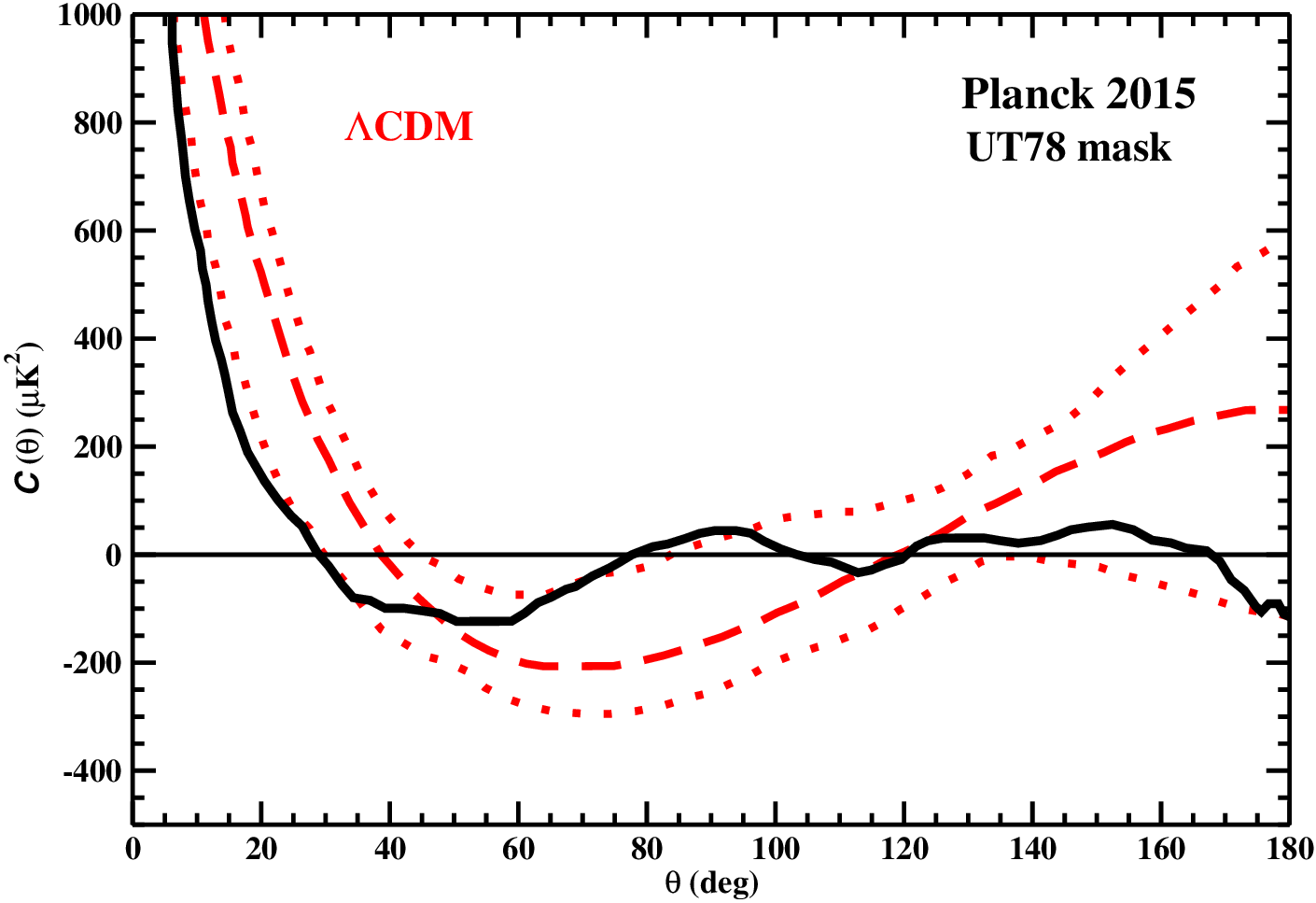}
\caption{\label{Fig1}  (Color online) The (black) continuous lines are  the two-point angular correlation function from the Planck 2013 (left panel)
and Planck 2015 (right panel) SMICA maps.
The (red) dashed lines are the best-fit $\Lambda$CDM model correlation functions together with the 68 \%  cosmic  variance
confidence intervals (red dotted lines). 
}
\vspace{0.3cm}
\end{figure}

The two-point temperature  correlation function is defined as the average product of two temperatures  measured in a fixed 
relative orientation on the sky:
\begin{equation}
\label{12}
\mathcal{C}(\theta)   \;  =  \;   \left <   \Delta T(\vec{n_1})   \Delta T(\vec{n_2})         \right > \;  \; , \; \; \vec{n_1} \cdot \vec{n_2} \; = \;  \cos \theta \; .
\end{equation}
Under the assumption of statistical isotropy, this correlation functions does not depend  on the particular position or orientation
on the sky  and,  thereby,  it depends only on the angle $\theta$.
The two-point angular correlation function is related to the angular power spectrum by:
\begin{equation}
\label{13}
\mathcal{C}(\theta)   \;  =  \;   T_0^2   \;  \sum_{\ell}  \frac{ 2 \ell+1}{4 \pi}  \;  C_{\ell} \;  P_{\ell}(\cos \theta)  \;  \; ,
\end{equation}
with  the related cosmic variance:
\begin{equation}
\label{14}
\left ( \sigma^{CV}(\theta) \right )^2    \;  =  \;   T_0^4   \;  \sum_{\ell}  \frac{ 2 \ell+1}{8 \pi^2}  \;  C_{\ell}^2  \;  P^2_{\ell}(\cos \theta)  \;  \; .
\end{equation}
Interestingly, the 2-point angular correlation function shows clear evidence of a lack of structure for large separation angles. 
The lack of correlations at large angular scales was clearly detected  by the observation of temperature
anisotropies by the  Cosmic Background Explorer (COBE)~\cite{Hinshaw:1996}, by
the Wilkinson Microwave Anisotropy Probe (WMAP)~\cite{Bennett:2011,Bennett:2013},  
by the first release of the Planck data (Planck 2013)~\cite{Ade:2014}
 and confirmed by the  Planck 2015~\cite{Ade:2016} and  Planck 2018 data~\cite{Akrami:2020}.
This lack of large-angle correlations in the observed microwave background temperature fluctuations probably is related to the lowness of the temperature quadrupole. In particular, there is a strong correlation between the low quadrupole and the lack of correlation in the two-point correlation 
function of the  CMB anisotropies~\cite{Gruppuso:2014a,Muir:2018}.
 Nevertheless, it is believed that it is a  different problem that, in principle, could  challenge the assumed fundamental
 prediction of gaussian random, statistically isotropic temperature fluctuations. Indeed,  this problem  has been subjected to several studies~\cite{Efstathiou:2004,Copi:2006,Copi:2007,Copi:2009,Copi:2010,Copi:2013,Gruppuso:2014b,
Copi:2015,Yoho:2015,Copi:2016a,Copi:2016b,Agullo:2021, Chiocchetta:2021}. \\
To illustrate the problem we display in Fig~\ref{Fig1}, (black) continuous lines,
the two-point angular  correlation function as observed with Planck.
More precisely,  Fig~\ref{Fig1}  (left panel),  adapted from Fig.~1 of Ref.~\cite{Copi:2015},  corresponds
to the full-sky  angular correlation function reported in  Ref.~\cite{Copi:2015} using the Planck 2013 data. 
On the other hand, in Fig~\ref{Fig1}  (right panel),  adapted from Ref.~\cite{Ade:2016},  it is shown  the
Planck 2015  measured angular  correlation function  with the UT78 mask.
The UT78 mask has a usable sky fraction of approximately 78  \% and it is the most conservative mask to omit foreground residuals.
The Planck team presented the analyses of the angular two-point correlation function at low resolution for
their four component separation methods (COMMANDER, NILC, SEVEM, SMICA). It turned out  that
the results of the Planck analyses by means of  the COMMANDER, SEVEM, NILC and SMICA maps 
fell on top of each other. Therefore, without loss in generality, in Fig~\ref{Fig1}  we restricted to
angular correlation function extracted with the SMICA map.  Moreover, it is useful to stress that
the observed angular correlation function by WMAP and Planck 2013, 2015 and 2018 are perfectly consistent each other.
The (red) dashed lines in  Fig~\ref{Fig1}  correspond to the expected two-point angular correlation based on comparison with
 $1000$ realisations of the best-fitting $\Lambda$CDM  cosmological model to the Planck data. 
We, also, display the 68 \% cosmic variance confidence interval (red dotted lines).
Comparing left and right panels in  Fig~\ref{Fig1}  one can see  how the temperature two-point correlation function
depends on the Galactic mask. 
Looking at Fig.~\ref{Fig1}, one sees that what is most striking is the difference between the best-fitting $\Lambda$CDM model and
the observed $\mathcal{C}(\theta)$. Indeed,  there is an evident suppression 
of power in the angular correlation function above about 60 degree. Moreover, this feature seems to be
a robust and statistically significant result. \\
There  are other puzzling aspects that it is worthwhile to mention. Firstly, the full-sky two-point angular correlation function
(left panel in Fig.~\ref{Fig1}) seems to vanish at three angular scales, $\theta_1 \simeq 31^\circ$,  $\theta_2 \simeq 96^\circ$ and  $\theta_3 \simeq 154^\circ$.
As a consequence, for angular scales larger than 150 degree, the observed two-point angular correlation function is negative,
while the expected one is positive. Even though at these large angular scales the best-fitted correlation function is affected by a
sizeable cosmic variance, it is quite difficult to image a physical mechanism able to account for such discrepancy. It should be  evident
that this problem is intimately connected with the observed suppression of the quadrupole temperature anisotropy.
In addition,  we feel that the another puzzling discrepancy resides on the fact that the expected two-point angular correlation function does
not track closely the observed   $\mathcal{C}(\theta)$ for small angular scale $\theta \lesssim 30^\circ$. Looking at Fig.~\ref{Fig1} we see
that the best-fitting two-point angular correlation function is systematically  higher than the  observed   $\mathcal{C}(\theta)$, even
though the difference in each angular bin is within one cosmic variance standard deviation. However, the cumulative effect of
the deviations can hardly be due to statistical fluctuations. In this regards, however, it should be mentioned that the values of
 $\mathcal{C}(\theta)$ in different angular bins are correlated, so that the sizeable deviation between the expected $\Lambda$CDM
 and the observed curve could be not so significant at it may appear.  On the other hand, from Fig.~\ref{Fig1} one infers that
 the main effects of the Galactic mask  is to shift the angular correlation function towards smaller angular separations together
 with a further reduction of the signal al large angular scales. Notwithstanding, also  the masked two-point angular correlation function 
 confirms the discrepancies  between predictions and observations.
 In any case, if one believes that  these anomalous features of the two-point angular correlation function cannot be ascribed to statistical
  fluctuations, then  the resolution requires to  identify the physics underlying the anomalies. 
\section{The Ellipsoidal Universe}
\label{s-3}
Let us, now, consider the two-point angular correlation function in the Ellipsoidal Universe model. According to Eq.~(\ref{9}) we have:
\begin{equation}
\label{15}
\mathcal{C}^{El}(\theta)   \;  =  \;   T_0^2   \;  \sum_{\ell}  \frac{ 2 \ell+1}{4 \pi}  \;  C^{El}_{\ell} \;  P_{\ell}(\cos \theta)  \;  \; ,
\end{equation}
To determine the coefficients   $C^{El}_{\ell}$ we should perform the best fits to the CMB anisotropy data within the Ellipsoidal Universe
model. Unfortunately, we do not yet have at our disposal the best-fitting   $C^{El}_{\ell}$. Nevertheless, we can reconstruct the correlation function
$\mathcal{C}^{El}(\theta)$ if we restrict to the angular region $\theta  \gtrsim 2^\circ$.
\begin{figure}[t]
\vspace{-1.0cm}
\includegraphics[width=0.5\textwidth,clip]{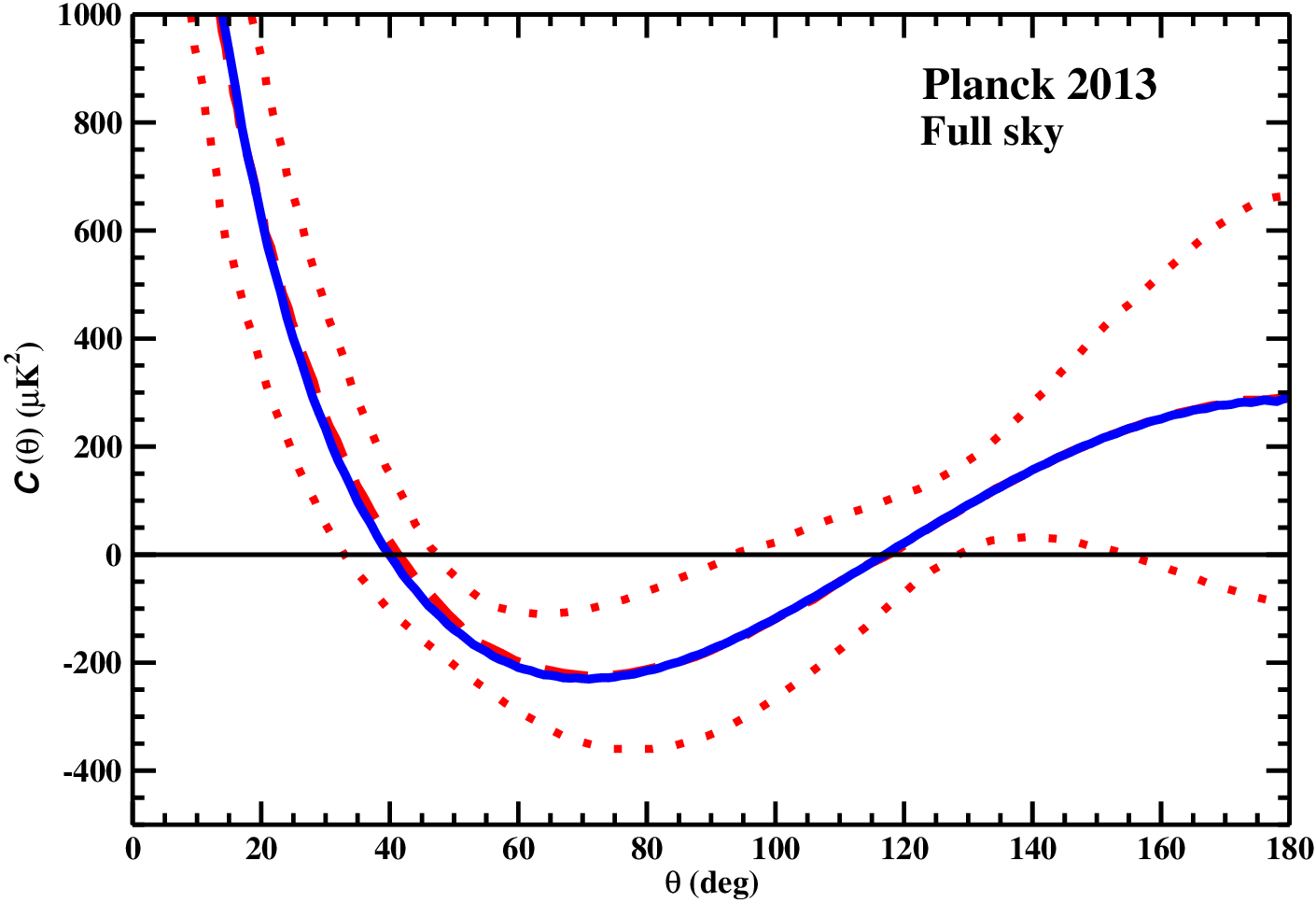}
\includegraphics[width=0.5\textwidth,clip]{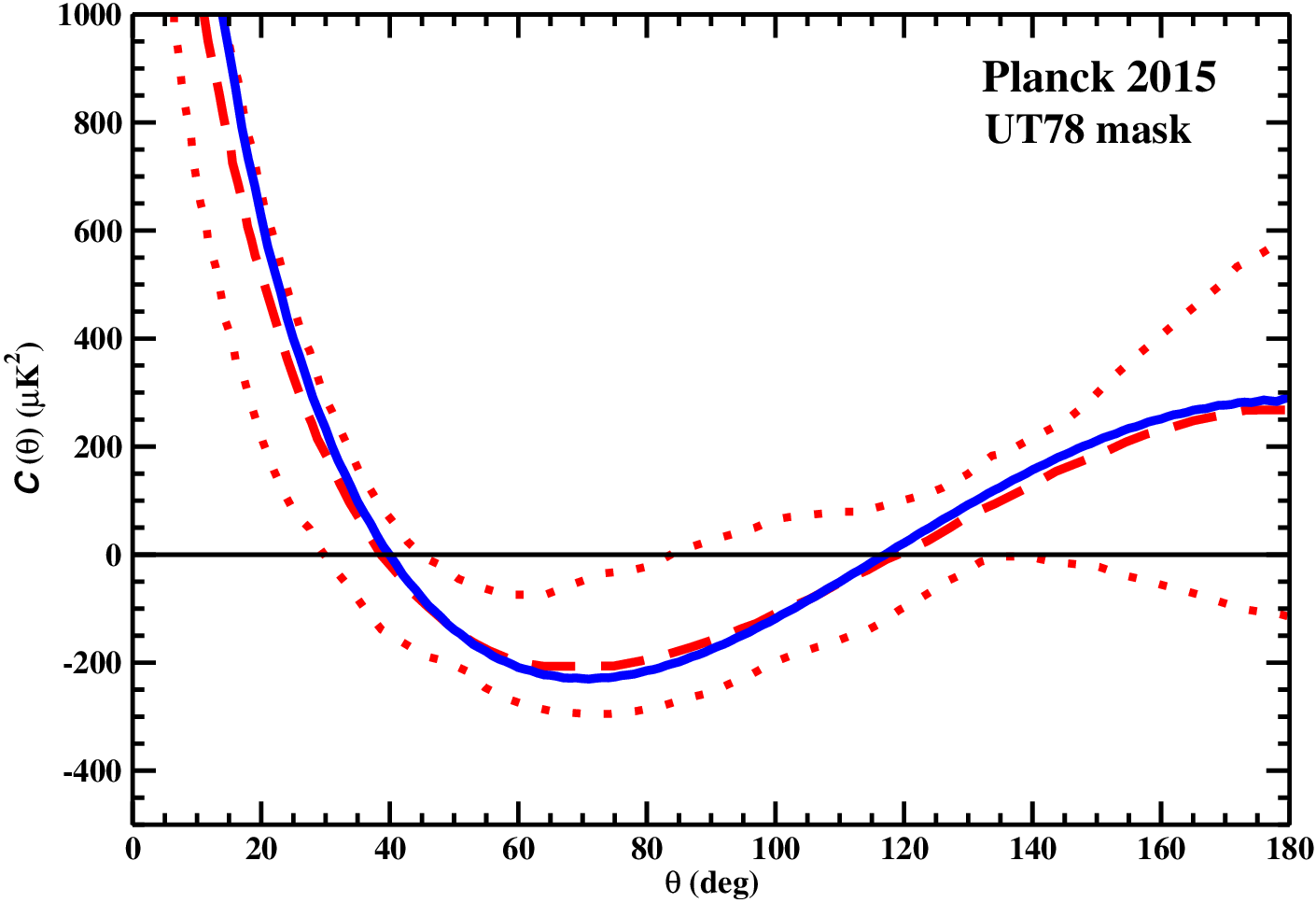}
\caption{\label{Fig2}  (Color online) Comparison of the best-fit $\Lambda$CDM model correlation functions (red dashed line)
to Eq.~(\ref{18}), blue continuous lines,  for  both the full-sky Planck 2013 data (left panel) and the Planck 2015 data with the   UT78 Galactic mask
(right panel).
}
\end{figure}
In fact, in the $\Lambda$CDM model the two-point temperature-temperature correlation function can be approximated as:
\begin{equation}
\label{16}
\mathcal{C}^{\Lambda CDM}(\theta)   \;  \simeq  \;   T_0^2   \; \;  \sum_{\ell=2}^{ \ell = \ell_{max}}  \; \; 
 \frac{ 2 \ell+1}{4 \pi}  \;  C^{\Lambda CDM}_{\ell} \;  P_{\ell}(\cos \theta)  
\end{equation}
where $\ell_{max} \ll 200$. Moreover, for $\ell \lesssim \ell_{max}$ the power spectrum coefficients can be estimated by the 
Sachs-Wolf effect (see, eg, Ref.~\cite{Dodelson:2021}):
\begin{equation}
\label{17}
\ell \; (\ell \; + \; 1)  \;  C^{\Lambda CDM}_{\ell} \;  \simeq \; \frac{8}{25} \; A_s \; \; , 
\end{equation}
where $A_s$ is the amplitude of the curvature power spectrum assuming  a scale-invariant  spectrum.
From Eqs.~(\ref{16}) and (\ref{17}) we get:
\begin{equation}
\label{18}
\mathcal{C}^{\Lambda CDM}(\theta)   \;  \simeq  \;   T_0^2  \; \frac{8}{100 \, \pi}  \; A_s \; \;  \sum_{\ell=2}^{ \ell = \ell_{max}}  \; \; 
 \frac{ 2 \ell+1}{\ell (\ell + 1)}  \;   P_{\ell}(\cos \theta)   \; \; .
\end{equation}
In Fig.~\ref{Fig2} we compare the two-point angular correlation function Eq.~(\ref{18}) to the best-fitting $\Lambda$CDM model correlation
function. We fixed:
\begin{equation}
\label{19}
 \ell_{max}  \; \simeq  \;  100  \; \; \; , \; \; \; A_s \; \simeq \; 3.0 \times \; 10^{-9} 
\end{equation}
such that $\mathcal{C}^{\Lambda CDM}(\theta)$ in Eq.~(\ref{18}) reproduces as closely as possible the best-fit $\Lambda$CDM correlation function.
Indeed, from Fig.~\ref{Fig2} we infer that the given approximations to evaluate the angular correlation function are quite adeguate to our
purposes for both the full-sky and masked Planck SMICA maps.
This allows us to estimate the angular correlation function for the Ellipsoidal Universe cosmological model.
In fact, according to our previous discussion and taking into account  Eq.~(\ref{10}),  we can write:  
\begin{equation}
\label{20}
C^{El}_{2} \; \simeq \; 0.26 \;    C^{\Lambda CDM}_{2}  \; \; , \; \;
 C^{El}_{\ell} \; \simeq \;  C^{\Lambda CDM}_{\ell} \; \; , \; \; \ell \; \ge \; 3 
\end{equation}
where we have taken into account that $(\Delta T_2)^2/(\Delta T_2^{\Lambda CDM})^2 \simeq 0.26$.
Accordingly, we have:
\begin{equation}
\label{21}
\mathcal{C}^{El}(\theta)   \;  \simeq  \;   T_0^2  \; \frac{8}{100 \, \pi}  \; A_s \;  \left (  0.26 \times \; \frac{5}{6} \;   P_{2}(\cos \theta) \; + \; 
  \sum_{\ell=3}^{ \ell = \ell_{max}}  \; \;   \frac{ 2 \ell+1}{\ell (\ell + 1)}  \;   P_{\ell}(\cos \theta)   \right ) \; \; .
\end{equation}
\begin{figure}
\vspace{-0.5cm}
\vspace{-1.0cm}
\includegraphics[width=0.5\textwidth,clip]{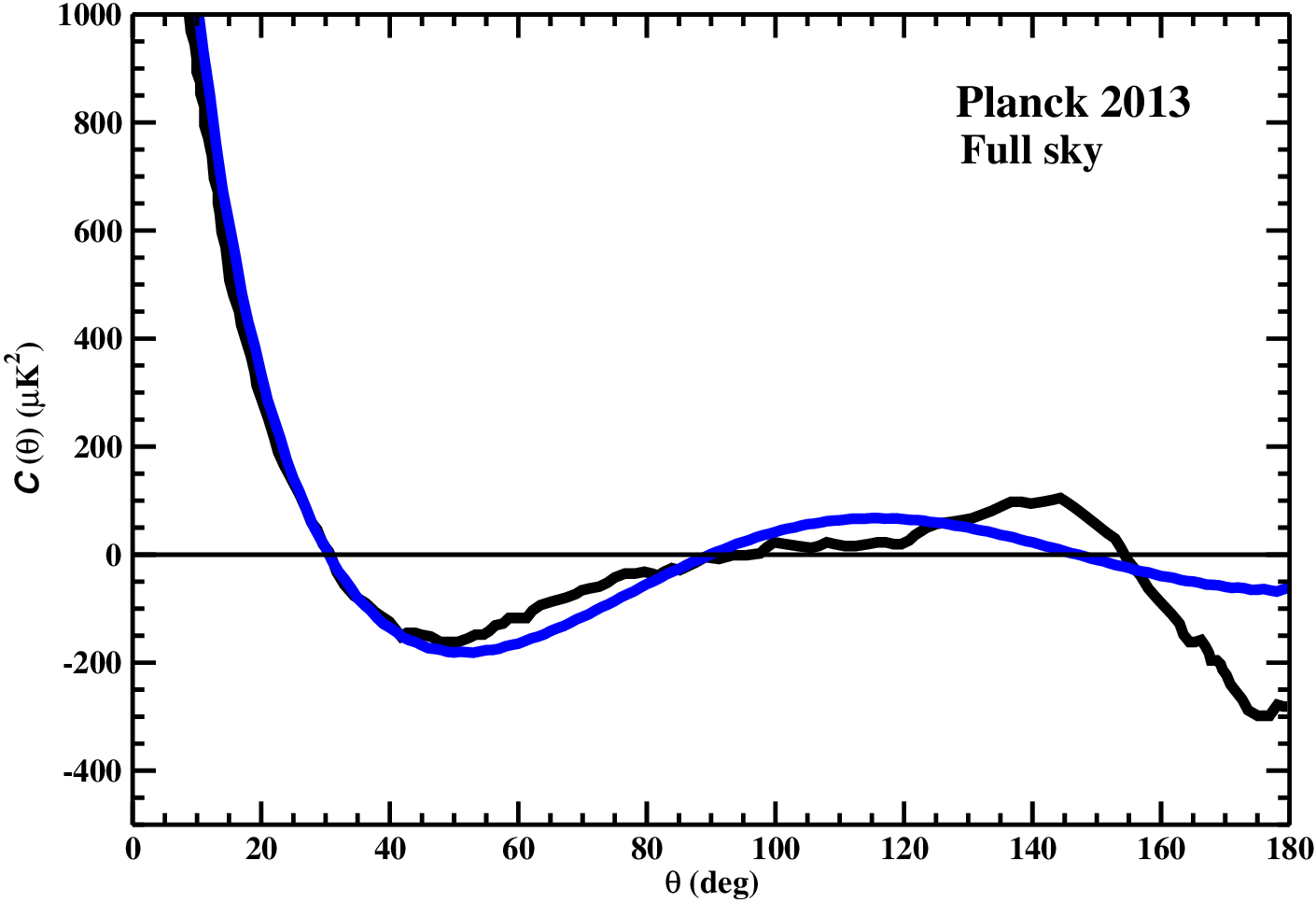}
\includegraphics[width=0.5\textwidth,clip]{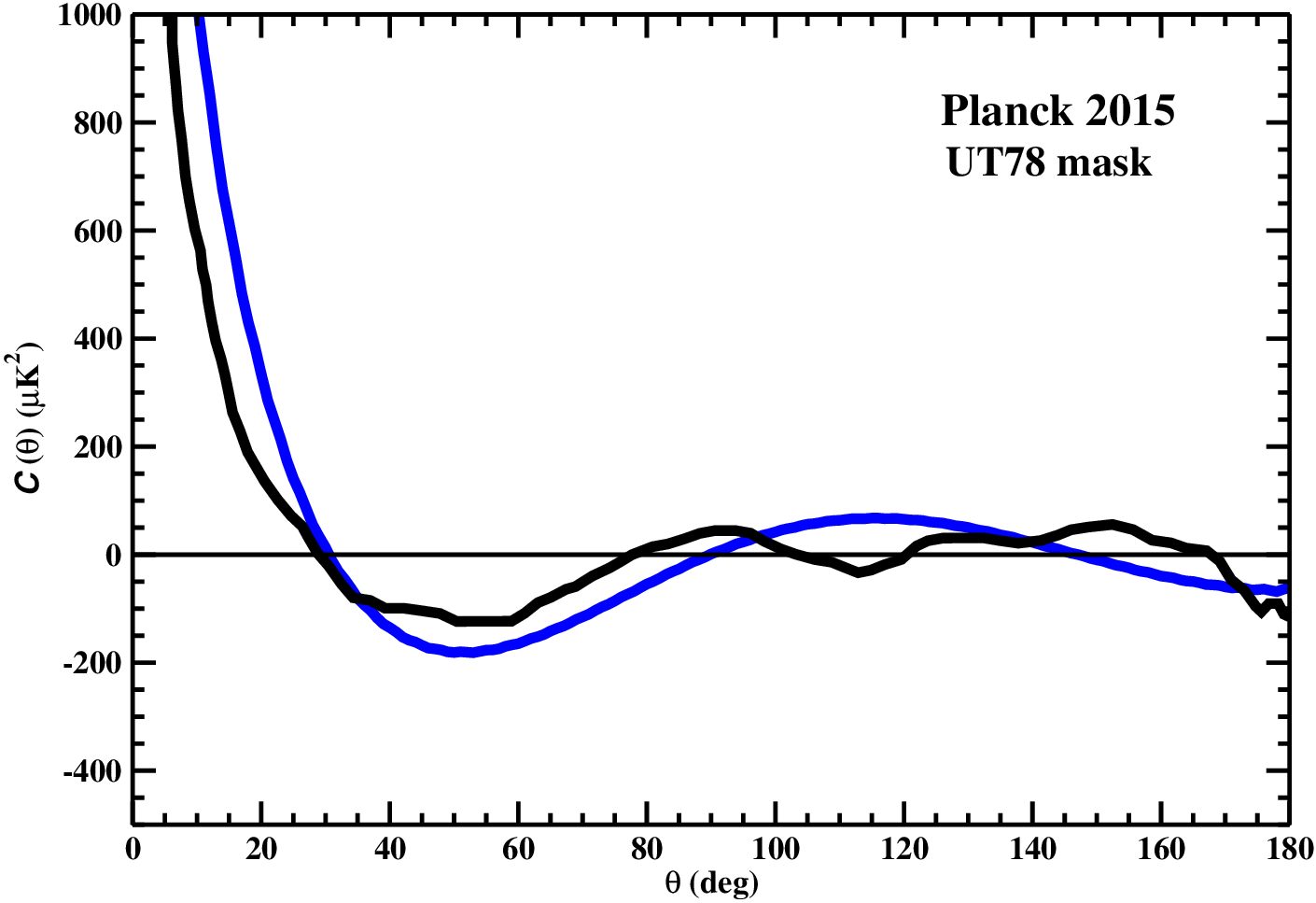}
\caption{\label{Fig3} 
(Color online) Comparison of the Ellipsoidal Universe model correlation function (blue continuous lines) as given by Eq.~(\ref{21})
to the  two-point angular correlation function from full-sky Planck 2013 (left panel) and Planck 2015  UT78 mask (right panel) 
SMICA maps (black continuous lines).
}
\end{figure}
The main results of the present paper are displayed in Fig.~\ref{Fig3} where we contrast $\mathcal{C}^{El}(\theta)$, Eq.~(\ref{21}), 
to the observed two-point angular correlation function from the Planck SMILCA maps. In the case of the full-sky angular correlation function
we see that the Ellipsoidal Universe model correlation function is able to trace closely the observed correlation function. There are 
some small deviations at very high angular separations that, however, are well within the cosmic variance uncertainties. In any case,
please note that our theoretical curve predicts correctly  a negative angular correlation function  for angular scales larger than 150 degree.
Even for  the masked angular correlation function the agreement between theoretical expectations and observations seems to
be satisfying. In this case there are some small deviations even at small angular scales that, as already discussed, can be ascribed
to the Galactic mask. \\
To quantify the lack of power on angular scales greater than 60$^0$  in Ref.~\cite{Spergel:2003} it was  introduced the parameter:
\begin{equation}
\label{22}
S_{1/2} \; = \; \int_1^{1/2} \;  [ \mathcal{C}(\theta) ]^2  \;  d (\cos  \theta) \; .
\end{equation}
It is useful to determine   this parameter  for both $\Lambda$CDM and Ellipsoidal
Universe models. To this end we have evaluated numerically the integrals in Eq.~(\ref{22})  by using Eqs.~(\ref{18})  and (\ref{21}). 
We  get:
\begin{equation}
\label{23}
S^{\Lambda CDM}_{1/2} \;  \simeq  \;  42527 \; (\mu K)^4 ,
\end{equation}
\begin{equation}
\label{24}
S^{El}_{1/2} \;  \simeq  \;  6837  \; (\mu K)^4 .
\end{equation}
For comparison we, also, display the value of the parameter S$_{1/2}$  based on the WMAP 5-year anisotropy measurements as reported in
 Table 1 of Ref.~\cite{Copi:2009}:
\begin{equation}
\label{25}
S_{1/2} \;  \simeq  \;  8833  \; (\mu K)^4 \; , \; \; WMAP \; \; 5-year \; .
\end{equation}
Comparing this last equation with  Eqs.~(\ref{23}) and (\ref{24}) one sees  that
the Ellipsoidal Universe cosmological model seems to be  in better agreement with observations with respect to the standard 
$\Lambda$CDM cosmological model. However, it should be keep in mind that the lack of correlations at large angular scales
in the correlation function is due almost entirely to the suppression of power in the quadrupole temperature anisotropy and that
the quadrupole $C_2$  is subject to a large  intrinsic uncertainty  given by the cosmic variance, $\Delta C_{2}  = \sqrt{\frac{2}{5}}  C_{2}$.
Actually, if we allow the quadrupole coefficient to vary in the interval ($C_2 - \Delta C_2, C_2 + \Delta C_2$), then
both the standard $\Lambda$ Cold Dark Matter cosmological model and the  Ellipsoidal Universe cosmological model
are consistent with the observed $S_{1/2}$, as given by  Eq.~(\ref{25}), at the 68 \% confidence level. \\
It should, now,  be evident that the Ellipsoidal Universe cosmological model compares rather well to the Planck observations. 
It is worthwhile to stress that to recover the anomalous features in the CMB angular correlation function  one
must admit that the temperature quadrupole suppression is a truly physical effect and not a mere statistical fluctuation.
In this respect, it should be mentioned that there are also other models that are able to explain the low quadrupole. For instance,
 a fast roll phase of the inflation preceding the slow roll phase is an explanation often considered in the literature~\cite{Gruppuso:2015}, or 
the introduction of a hard lower cutoff in the primordial power  spectrum~\cite{Sanchis-Lozano:2022}.
\section{Summary and Conclusions}
\label{s-4}
The latest  results on the CMB anisotropies by the Planck Collaboration are confirming the
standard $\Lambda$ Cold Dark Matter cosmological model with an exquisite  level of accuracy.
Nevertheless, at large angular scales there are still anomalous features in CMB anisotropies. Actually,
the most evident discrepancy resides in the quadrupole temperature correlation.
It is  conventional wisdom to believe that this quadrupole anomaly is due to a statistical fluctuation.
However,  there is also a persistent anomaly   in the temperature two-point angular correlation function computed 
as an average over the full sky.  We have shown that, if we consider the quadrupole suppression
a truly physical effect, then we can account for the persistent  lack of correlations at large angular scales
in the two-point temperature angular correlation function.  This last point implies
that the standard cosmological model  necessitates  some changes.
Remarkably, the Ellipsoidal Universe cosmological model, advanced several years ago
to account  for the CMB quadrupole anomaly,  constitutes a  viable alternative to
the standard cosmological model. In fact,  if one assumes that the large-scale spatial geometry of our Universe
is slightly anisotropic, then the quadrupole amplitude can be drastically reduced without
affecting higher multipoles of the angular power spectrum of the temperature anisotropies.
At the same time, we showed in the present paper that the Ellipsoidal Universe two-point angular correlation function
compares reasonable well to observations.
On the other hand, at variance of the standard cosmological model,
 it is known since long time that anisotropic cosmological model could induce sizeable
large-scale CMB polarisation~\cite{Rees:1968,Anile:1974,Basko:1980,Negroponte:1980}. 
Indeed, we already argued  in Refs.~\cite{Cea:2014,Cea:2020} that in the Ellipsoidal Universe
model there is a sizeable polarisation signal at scales $\ell  \lesssim  10$. Moreover, we showed that the
quadrupole TE and EE correlations in the Ellipsoidal Universe are in reasonable agreement
with the Planck 2018 data. Finally, quite recently, we suggested~\cite{Cea:2022} that the Ellipsoidal Universe model should also
alleviate the tensions on the Hubble constant $H_0$ and the cosmological parameter $S_8$. \\
In conclusion, we have shown that the Ellipsoidal Universe cosmological model allows to explain several anomalous features
in the CMB temperature anisotropies.  Our results are suggesting that   the Ellipsoidal Universe cosmological model 
is not only a viable alternative to the $\Lambda$CDM cosmological model, but also it seems to compare observations
slightly better than  the standard cosmological model. Finally,  we would like to conclude the present paper by
stressing that, due to  the low statistical significance, by using only the CMB temperature anisotropies
 one cannot distinguish  the $\Lambda$CDM cosmological model from eventual extensions. One way  to overcome this problem 
 might be to consider the large-scale polarisation.  Unfortunately the polarisation Planck data at low $\ell$  are not signal-dominated as 
 in temperature. The future CMB experiments sensitive to the very low multipoles of the CMB polarisation, such as the LiteBIRD satellite,
 may provide us important information  about it.  Indeed, LiteBIRD represents the fourth generation of satellites dedicated to 
 the CMB following its predecessors COBE, WMAP and Planck and it will be the first  completely dedicated to the CMB polarisation. 
 Actually,  LiteBIRD's primary goal is to map the   microwave sky in polarisation on large angular  scales with an unprecedented 
 sensitivity~\cite{LiteBIRD:2022}.

\end{document}